\documentclass[twocolumn,showpacs,preprintnumbers,amsmath,amssymb]{revtex4-1}%
\usepackage{graphicx}
\usepackage{dcolumn}
\usepackage{bm}
\usepackage{amsmath}
\usepackage{amsfonts}
\usepackage{amssymb}%
\begin{document}
%\preprint{APS/123-QED}
\title{Topological matter with collective encoding and Rydberg blockade}
\author{Anne E. B. Nielsen and Klaus M\o lmer}
\affiliation{Lundbeck Foundation Theoretical Center for
Quantum System Research, Department of Physics and Astronomy,
Aarhus University, DK-8000 \AA rhus C, Denmark}
\date{\today}

\begin{abstract}
We propose to use a permutation symmetric sample of multi-level atoms to simulate the properties of topologically ordered states. The Rydberg blockade interaction is used to prepare states of the sample which are equivalent to resonating valence bond states, Laughlin states, and string-net condensates and to create and study the properties of their quasi-particle-like fundamental excitations.
\end{abstract}


\maketitle

\section{Introduction}

Quantum correlations between particles constitute an important topic in quantum information research. Universal schemes for quantum computing allow the formation of arbitrarily correlated states from simple product states of two-level quantum systems, and via sequential application of one- and two-particle interactions, a wide variety of many-body interactions may be effectively simulated \cite{Feynman,Lloyd,Cirac_simulator,Lewenstein}. Particular interest has been devoted to topological matter which shows correlations but no local order parameter \cite{kitaev,pachos,whaley}.  The Laughlin wave-function describing the fractional quantum Hall effect is an example of a topological state of correlated electrons, while high-$T_c$ superconductivity has been proposed to originate from the resonating valence bond (RVB) state, which is a superposition of different spin singlet configurations.

While demonstration of topological matter in natural physical materials is so far restricted to the quantum Hall effect, a number of proposals exist to synthesize states of matter, which display the desired correlations. One may distinguish two kinds of implementation: i) synthesis of states of many particles which are formally equivalent to topological matter candidates, ii) synthesis of effective Hamiltonian interactions, which have the desired topological states as ground states or eigenstates. The first kind of implementation may be used to verify by measurement the value of certain correlations and effects of certain operations on the system, while the second kind may be used to illustrate phase transition dynamics, dynamical response, and robustness to external perturbations.

Topological states on a two dimensional array of spin $1/2$ particles or qubits, are interesting candidates for quantum memories stored in collective states of many particles. Quantum information is stored in states with large topological degeneracy, but with an excitation energy gap which protects against local disturbances \cite{kitaev,pachos,whaley,nayak}. Even without interactions, specifically ordered states offer error protection, if one is able to probe the multi-particle constraints and thus identify  single particle errors when they occur in quantum computing \cite{DiVincenzo2009} and communication \cite{Hollenberg} and suitably correct them.

Periodic arrangements of atoms in optical lattices physically display the superfluid-insulator dynamics of the Hubbard Hamiltonian \cite{Jaksch,Greiner}, and lattices with a single trapped atom per lattice site have been proposed as simulators of Ising and Heisenberg interacting spin models \cite{IsingAK}, and more elaborate lattice gauge theories \cite{lattice-gauge,weimer}, where the geometry of the potential wells in an optical lattice are here in immediate correspondence with the mathematical models of particles or spins on a square, triangular or honeycomb lattice.

In \cite{paredes}, Paredes and Bloch offer an excellent review of the properties of a variety of topological states together with proposals for implementation of these states with neutral atoms trapped in optical lattice potentials. In particular, they focus on a single plaquette, \textit{i.e.}, an arrangement of four spin $1/2$ particles at the corners of a square as illustrated in Fig.~\ref{one}(a). If one represents the spin up (down) states by filled (empty) sites of hard core bosons, one obtains the state illustrated in Fig.~\ref{one}(b).

Tunnelling among wells combined with controlled ground state collisions \cite{paredes} and long range dipole interaction between atoms excited to high lying Rydberg states \cite{lattice-gauge} have been proposed to engineer both unitary and dissipative evolution leading to formation of topological states in the lattice system.

In this work, we will show how Rydberg blockade in small atomic ensembles in conjunction with qubit encoding in the collective atomic population in different internal atomic states \cite{brion} can be used to simulate simple instances of topological matter. In particular, we will show that a few hundred atoms with five stable internal states can be used to simulate Laughlin-like states, RVB-states, and flux and charge quasi-particle excited states of a simple plaquette system. Such states provide an interesting stepping-stone for experimental demonstrations and investigations of topological effects. The collective qubit encoding in atomic ensembles is formally equivalent to photonic quantum computing schemes. Multi-photon states have been produced with correlations mimicking the spin correlations of simple models of topological matter \cite{pan,weinfurter}. The photonic implementations use squeezed light sources, and detector coincidence signals verify the values of quantum correlations in a heralded (and destroyed) component of the light. In contrast our atomic scheme generates the states on demand and permits sequences of subsequent measurement and interaction operations to be carried out on the system.

In Sec.~\ref{plaquette}, we discuss the main features of the plaquette model that we will implement and study in this paper. In Sec.~\ref{collective}, we review the Rydberg blockade mechanism and we discuss and analyze the schemes for collective qubit encoding and manipulation needed to simulate topological states in small atomic ensembles. In Sec.~\ref{manipulation}, we present effective protocols to engineer particular topological quantum states in collective atomic degrees of freedom, and we perform a variety of operations on these states. Section \ref{conclusion} concludes the paper.

\section{The plaquette}\label{plaquette}

\begin{figure}
\includegraphics[width=0.6\columnwidth]{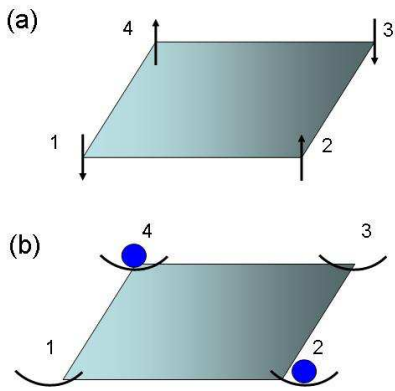}
\caption{(Color online) (a) A plaquette of four spins in an optical lattice. (b) Equivalent plaquette of hard core bosons with a filled (empty) site representing the spin up (down) states in part (a) of the figure.}\label{one}
\end{figure}

A simple instance of topological matter can be simulated by four spin 1/2 particles arranged in the corners of a square and enumerated 1-4. Figure~\ref{one}(a) shows a classical spin configuration with spins 1 and 3 in spin down states and spins 2 and 4 in spin up states. Spin singlet states of any short range pair of spins, \textit{i.e.}, spins on the same side of the plaquette such as spins 1 and 2, can be defined, and the resonating valence bond (RVB) states are superpositions of states where the spin singlets occur for the pairs (1,2) and (3,4) and for the pairs (1,4) and (2,3). These RVB states are examples of topological spin liquids.

In Fig.~\ref{one}(b), the spin down particle is replaced by an empty site, and the spin up particle is replaced by a site occupied by exactly one particle. The Hilbert space of four spins with a vanishing total $z$-component is thus equivalent to the space of two hard core bosons distributed among four spatial locations with at most one boson per site. The state shown is thus in one-to-one correspondence with the state shown in part (a) of the figure, but the quantum degree of freedom is here associated with the occupation of the sites, \textit{i.e.}, by the spatial location of physical particles, rather than by the internal spin degree of freedom of space fixed particles. Note that it is a natural property of this representation that a single particle occupying an odd superposition state of two lattice locations is equivalent to an entangled spin singlet state of the corresponding two spin $1/2$ particles.

In \cite{paredes}, the analogy of the hard core boson occupation degree of freedom and the spin 1/2 internal states is discussed in detail, and it is proposed how laser double well potentials along two spatial directions can be used to define the plaquette and how tuning of the on-site interaction and the tunneling of particles may be used to engineer at will a variety of different states in both the spin and in the hard core boson representation. It is for example possible to allow tunneling with amplitude $t_x$ along the 1-2 and 3-4 bonds, causing a super exchange interaction
$H_S= J_x (\hat{P}_{12} +\hat{P}_{34})$, where $\hat{P}_{ij}$ is the projection operator on the spin singlet state of spins $i$ and $j$, $J_x=4t_x^2/U$ with $U$ the on-site interaction energy. The product state of $(1,2)$ and $(3,4)$ spin singlets is then gapped from all other states of the four spins, and by lowering the potential barrier in the orthogonal direction and thus increasing the tunneling amplitude, $t_y$, along the 1-4 and 2-3 bonds, one provides a Hamiltonian
\begin{equation}
H_S= J_x (\hat{P}_{12} +\hat{P}_{34}) + J_y (\hat{P}_{14} +\hat{P}_{23}),
\end{equation}
which adiabatically connects the product state of two simple spin singlets with one of the RVB states. In the hard core boson representation (removing the spin down particles) the RVB state of the remaining two spin up particles is also equivalent to the Laughlin state, and in \cite{paredes} it is proposed how further use of external inhomogeneous perturbations can engineer quasi-particle excited states and simulate their anyonic behavior associated with the excitations being propagated around all corners of the plaquette.

\section{Rydberg blockade and collective encoding of qubits}\label{collective}

\begin{figure}
\includegraphics[width=0.6\columnwidth,clip]{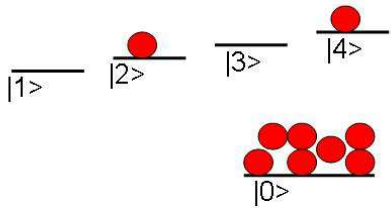}
\caption{(Color online) Collective encoding of a four qubit register with zero or unity collective occupation in four different internal atomic states. The state shown is equivalent to the hard core boson plaquette illustrated in Fig.~\ref{one}(b).}\label{two}
\end{figure}

Instead of coding memory bits into separate particles, it has been proposed to approach scalability of quantum information processing by making use of the rich internal level structure found in many atomic and molecular quantum systems \cite{brion,tordrup}. In these proposals it is suggested to use ensembles of identical particles and to encode quantum information in symmetric states of the ensemble with different values of the
collective population of the different internal atomic or molecular states. To be more specific, an arbitrary $N$-bit register state $\left\vert b_{1}b_{2}\ldots b_{N}\right\rangle$, with $b_{i}=0,1$, is encoded
in the collective symmetric state of $K$ identical quantum systems with $\left( N+1\right)$ internal levels $\left\{ \left\vert i\right\rangle ,i=0,\ldots,N\right\}$ if one lets precisely $b_{i}$ particles populate the internal state $\left\vert i\right\rangle $, see Fig.~\ref{two}, while $K-\sum_{i=1}^{N}b_{i}$ particles remain in the so-called reservoir state $\left\vert 0\right\rangle$. Values of $b_i$ that are larger than unity are not compatible with the binary logic of qubits, but access to an extra quantum level associated with each qubit can lead to very efficient operations, as we shall see in the following.

The system may be initialized with all atoms prepared in the state $|0\rangle$, e.g., by optical pumping, and provided the lasers and other means used to excite and control the ensemble act identically on all atoms, the state will retain its permutation symmetry for all times. Note that for atoms at rest, e.g., trapped in an optical lattice potential, phase factors $e^{i{\bf k}\cdot{\bf r}_j}$ arising from travelling wave laser fields which excite the atoms at their different spatial locations can be absorbed in a redefinition of the atomic internal state coherences. To force the system to remain in the subspace of meaningful quantum register states we propose to apply the Rydberg blockade mechanism \cite{jaksch,lukin}, which uses the strong long-range interaction between highly excited atoms and the associated energy shifts to prevent atoms in the vicinity of an already excited neighbor from being transferred into a Rydberg state. For a review of quantum information processing ideas with Rydberg excited atoms and relevant physical parameters, see \cite{Rydbergreview}.

In an ensemble of atoms, which are all within reach of these strong interactions, extending out to 10~$\mu$m separation, a resonant laser process towards a definite Rydberg state may produce an excited superposition state
\begin{equation}
|\overline{r}\rangle = \frac{1}{\sqrt{K}} \sum_{j=1}^K |0_10_2 \ldots 0_{j-1} r_j 0_{j+1} \ldots 0_K\rangle
\end{equation}
in which all atoms are excited with the same probability amplitude, but the laser field resonant with a single Rydberg excitation is detuned for states with two or more Rydberg excited atoms because of the interaction energy between excited atoms. Taking advantage of this mechanism one can drive precisely one atom from the initially macroscopically populated state $|0\rangle$, via a Rydberg state $|r\rangle$, into an initially empty internal state, and thus prepare a collective state with precisely one atom or no atom in any of the register states $|i\rangle$.

The coherent excitation towards the Rydberg state may be terminated so that a superposition of a state with zero and one Rydberg atom is produced, and in this way, we can prepare quantum superposition states of the qubit register. The excitation to a Rydberg level from one internal state can also prevent excitation from another internal state, and in this way two-qubit entangling operations can be implemented through the same mechanism \cite{brion}.
In fact, the universal power of the collective encoding quantum computing allows us to synthesize any state and to implement any unitary transformation within the atomic ensemble.

It is important to point out, that the operations mentioned are carried out by collective addressing, and the exciting laser pulses only need to be adjusted to the transition frequencies, the coupling strengths and the polarization selection rules of the atomic transitions. A practical implementation in alkali atoms may thus apply a homogeneous magnetic field, which splits the different Zeeman sublevels of the atomic hyperfine ground state structure, and thus the excitation towards the Rydberg states and lower lying optically excited states can be made state specific via the transition frequencies \cite{brion}.

\subsection{Entanglement generation with and without interactions}

We noted already that the universal quantum computing capacity of the collective coding scheme and the Rydberg blockade mechanism permit the generation of any state of the system. In this subsection we wish to point out that some states are actually more easily synthesized in an alternative manner than by processing a sequence of binary logic quantum gates. As mentioned in Sec.~\ref{plaquette}, a single hard core boson in an odd superposition state of two spatial locations is equivalent to a spin singlet state, \textit{i.e.}, an entangled state of two localized spin $1/2$ particles. With our collective encoding in Fig.~\ref{two}, we similarly observe that a logical state like $|1000\rangle$ with one atom in state 1, and no atoms in the other register states, can be transformed into any superposition $c_1 |1000\rangle + c_2 |0100\rangle + c_3 |0010\rangle + c_4 |0001\rangle$ by simply applying (Raman) laser fields onto the entire ensemble that transfer non-interacting atoms from the internal state $|1\rangle$ into the superposition $\sum_i c_i |i\rangle$.

It may seem puzzling that without use of any interaction between the atoms, it is possible to turn a classical register state into a quantum state corresponding to  entangled qubits.  In registers with individual quantum systems encoding of the different qubits such operations require direct or indirect interactions between the particles. But we recall that the ``classical'' register state $|1000\rangle$ is indeed a multi-particle entangled W-state of the atomic ensemble, where every atom has the same amplitude to populate the internal state $|1\rangle$. We observe that this atom-atom entanglement can be freely transformed into qubit-qubit entanglement within the register subspace of only a single bit value of unity. While we need the Rydberg blockade to manipulate the collective population between zero and unity in the register states, in the conventional encoding, one-bit unitary gates do not need any interaction.

For our purpose, it follows from the above that pairs of spin singlets of spins $(1,2)$ and $(3,4)$ are readily obtained from the state shown in Fig.~\ref{two} by simple Raman processes, while the superposition of $(1,2),(3,4)$ and $(1,4),(2,3)$ singlet pairs in the RVB states could be non-trivial two-particle states and hence deserve special attention. We will now present a general criterion for which two-atom states can be transformed into each other by simple Raman processes, i.e., by simple linear transformations on the single atom level structure.

Symmetric collective states of $K \gg 1$ atoms with two atoms in the register states may be conveniently represented as
\begin{equation}
|\psi_i\rangle=\sum_{i=1}^N\sum_{j=1}^N\hat{a}_i^\dag c_{ij}\hat{a}_j^\dag|00\ldots0\rangle,
\end{equation}
where the $\hat{a}_i^\dag$ operators are bosonic creation operators of an atom in level $i$. The bosonic property does not rely on the use of bosonic atoms in the ensemble, but merely reflects the permutation symmetry among the atoms.  $c$ is a matrix of coefficients determining the state, and for uniqueness, we require $c_{ij}=c_{ji}$. A Raman transition between levels $i$ and $j$ (where $i$ and $j$ may refer to the same level) is described by a single atom Hamiltonian $H_1 = \hbar (g|j\rangle\langle i|+g^*|i\rangle\langle j|)$ which gives rise to the multi-atom, second quantized Hamiltonian $H=\hbar (g\hat{a}_j^\dag\hat{a}_i+g^*\hat{a}_i^\dag\hat{a}_j)$. In the Heisenberg picture, this Hamiltonian causes a linear transformation among the bosonic creation operators, and by choosing suitable combinations of Raman transitions, it is possible to implement any linear transformation of the form $\hat{a}_i^\dag\rightarrow \sum_jU_{ij}\hat{a}_j^\dag$, where $U$ is a unitary matrix (the unitarity preserves the bosonic commutator relations $[\hat{a}_i,\hat{a}_j^\dag]=\delta_{ij}$). This transformation transforms the ensemble state $|\psi_i\rangle$ into
\begin{equation}
|\psi_f\rangle=\sum_{i=1}^N\sum_{j=1}^N\sum_{k=1}^N\sum_{l=1}^N \hat{a}_l^\dag U_{il} c_{ij}U_{jk}\hat{a}_k^\dag|00\ldots0\rangle.
\end{equation}
We can thus transform a symmetric ensemble state $|\psi_i\rangle$ with expansion coefficients $c$ into a state with coefficients $\tilde{c}$ without interatomic interactions if and only if we can find a unitary matrix $U$ so that
\begin{equation}\label{condition}
\tilde{c}=U^TcU,
\end{equation}
where $T$ denotes the transpose.

According to the Takagi decomposition \cite{Takagi}, it is always possible to write an $n\times n$ complex symmetric matrix $A$ as $A=V\Sigma V^T$, where $V$ is a unitary matrix and $\Sigma$ is a real nonnegative diagonal matrix. The diagonal elements of $\Sigma$ are the positive square roots of the eigenvalues of $AA^*$ and the column vectors of $V$ are a set of orthonormal eigenvectors of $AA^*$. It thus follows that two states with coefficients $c$ and $\tilde{c}$ can be transformed into each other by use of independent atom (Raman) transitions if and only if the Hermitian matrices $cc^*$ and $\tilde{c}\tilde{c}^*$ have identical eigenvalues. (For $c=V\Sigma V^T$, $\tilde{c}=\tilde{V}\tilde{\Sigma}\tilde{V}^T$, and $\Sigma=\tilde{\Sigma}$, the solution of \eqref{condition} takes the form $U=V^*\tilde{V}^T$.)

This simple analysis is useful to assess which gate implementation of quantum algorithms can be most easily carried out in the collective coding scheme, e.g., on remote, non-interacting particles. In the next section we will apply the criterion \eqref{condition} and find simple means to produce different plaquette states.

\section{Preparation and manipulation of topological states}\label{manipulation}

\subsection{Hard core boson equivalence of resonance valence bond states}

As specific examples, let us assume $N=4$ and consider the hard core boson states equivalent with spin singlet superposition states by the identification of the spin singlet creation operator $s_{ij}^\dag=(\hat{a}_i^\dag-\hat{a}_j^\dag)/\sqrt{2}$,
\begin{multline}
|\Phi_-\rangle=(s_{12}^\dag s_{43}^\dag-s_{14}^\dag s_{23}^\dag)|0000\rangle =(|1001\rangle+|0110\rangle\\
-|0011\rangle-|1100\rangle)/2
\end{multline}
and
\begin{multline}
|\Phi_+\rangle=(s_{12}^\dag s_{43}^\dag+s_{14}^\dag s_{23}^\dag)|0000\rangle/\sqrt{3}
=(|1001\rangle+|0110\rangle\\
+|1100\rangle+|0011\rangle-2|0101\rangle-2|1010\rangle)/\sqrt{12}.
\end{multline}
These are the resonating valence bond states of the simple plaquette described in \cite{paredes}. The coefficient matrix for $|\Phi_-\rangle$ is determined as
\begin{equation}\label{ctminus}
\tilde{c}=\frac{1}{4}\left[\begin{array}{cccc}0&-1&0&1\\
-1&0&1&0\\
0&1&0&-1\\
1&0&-1&0\end{array}\right],
\end{equation}
and the eigenvalues of $\tilde{c}\tilde{c}^*$ read $0$, $0$, $1/4$ and $1/4$.  $|\psi_i\rangle=|1100\rangle$ is a simple state to prepare by transfer of two atoms via the Rydberg state to states $|1\rangle$ and $|2\rangle$, and with the coefficient matrix
\begin{equation}\label{cini}
c=\frac{1}{2}\left[\begin{array}{cccc}0&1&0&0\\
1&0&0&0\\
0&0&0&0\\
0&0&0&0\end{array}\right],
\end{equation}
we find that $cc^*$ has the same eigenvalues, $0$, $0$, $1/4$ and $1/4$. This implies that the state  $|\Phi_-\rangle$ can be prepared from $|\psi_i\rangle=|1100\rangle$ by a simple transformation, and we compute the corresponding unitary single atom evolution operator,
\begin{equation}
U=\frac{1}{\sqrt{2}}\left[\begin{array}{cccc}-1&0&1&0\\
0&1&0&-1\\
1&0&1&0\\
0&1&0&1\end{array}\right].
\end{equation}
This matrix represents a $\pi/2$ pulse between levels 1 and 3 and a $\pi/2$ pulse between levels 2 and 4 with suitably chosen phases, which appears because $|\Phi_-\rangle$ is, indeed, a direct product of a singlet between spins 1 and 3 and a singlet between spins 2 and 4 \cite{paredes}.

We next consider the state $|\Phi_+\rangle$, which has the coefficient matrix
\begin{equation}
\tilde{c}=\frac{1}{4\sqrt{3}}\left[\begin{array}{cccc}0&1&-2&1\\
1&0&1&-2\\
-2&1&0&1\\
1&-2&1&0\end{array}\right]
\end{equation}
and the eigenvalues of $\tilde{c}\tilde{c}^*$ read $0$, $1/12$, $1/12$, and $1/3$. For ``classical'' register states with only one term, the eigenvalues are always either $0$, $0$, $1/4$, and $1/4$ or $0$, $0$, $0$, and $1$, and it is thus impossible to prepare $|\Phi_+\rangle$ from such simple states without using the Rydberg blockade (or another interaction between the atoms). We can still, however, use our mapping criterion, and, e.g., verify that the state $\sqrt{2}c_{11}|2000\rangle+(c_{23}+c_{32})|0110\rangle$ (where $|2000\rangle$ is the state with two atoms in level $|1\rangle$) leads to the eigenvalues $0$, $|c_{11}|^2$, $|c_{23}|^2$, and $|c_{32}|^2$. Choosing $c_{11}=1/\sqrt{3}$ and $c_{23}=c_{32}=1/(2\sqrt{3})$, we thus find the same eigenvalues as for $|\Phi_+\rangle$.
This implies, that we can find the single particle unitary transformation by which these two states are converted into each other, and it reads
\begin{equation}
U=\frac{1}{2}\left[\begin{array}{cccc}i&-i&i&-i\\
-i&1&i&-1\\
i&1&-i&-1\\
1&1&1&1\end{array}\right].
\end{equation}

\begin{figure*}
\includegraphics[width=\textwidth]{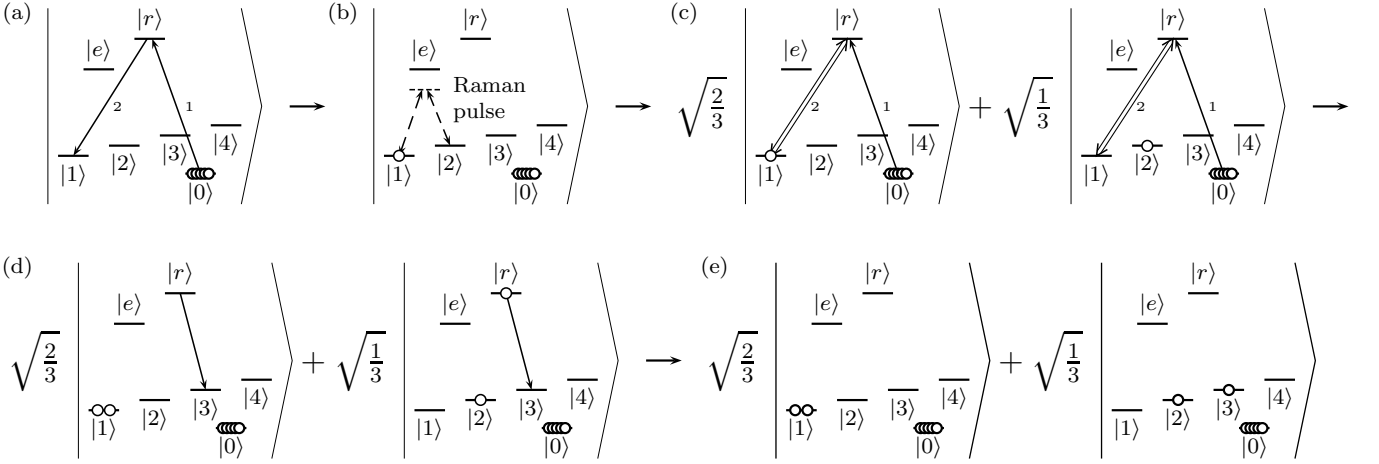}
\caption{Preparation of the state $\sqrt{2/3}|2000\rangle+\sqrt{1/3}|0110\rangle$ by use of a single Rydberg level. Full arrows are $\pi$ pulses, dashed arrows are Raman pulses, doublearrows are composite pulses, and $|e\rangle$ is an excited level. (a) Starting from the register state $|0000\rangle$, we apply two $\pi$ pulses in the indicated order to move precisely one atom from $|0\rangle$ to $|1\rangle$ as explained in Sec.~\ref{collective}. (b) We apply a Raman pulse to obtain the state $\sqrt{2/3}|1000\rangle+\sqrt{1/3}|0100\rangle$. (c) We first transfer one atom from $|0\rangle$ to $|r\rangle$ and then apply a composite pulse between levels $|1\rangle$ and $|r\rangle$, which transforms a state with one atom in $|1\rangle$ and one atom in $|r\rangle$ into a state with two atoms in $|1\rangle$ and no atoms in $|r\rangle$ and leaves a state with no atoms in $|1\rangle$ and one atom in $|r\rangle$ unchanged (see text). (d) We apply a $\pi$ pulse to interchange the populations in $|r\rangle$ and $|3\rangle$, which leads to the desired state (e).}\label{phiplus}
\end{figure*}

The state $\sqrt{2/3}|2000\rangle+\sqrt{1/3}|0110\rangle$ can be prepared as illustrated in Fig.~\ref{phiplus}. In part (c) of the figure, we use a composite pulse sequence consisting of three pulses with constant amplitude and phase, which was proposed in \cite{brionerror}. The ability to control the transfer of the atom in the Rydberg level into $|1\rangle$ dependent on whether $|1\rangle$ is initially populated by one or zero atoms ((c) to (d) in the figure) stems from the fact that the transition between a ground state level and a Rydberg level, due to the blockade effect, acts as a two-state transition with a Rabi frequency proportional to the square root of the total number of atoms occupying the two levels. The Rabi frequency on the $|1\rangle$ to $|r\rangle$ transition is thus a factor of $\sqrt{2}$ larger in the former case than in the latter case. We also investigated whether the transformation achieved with the composite pulse sequence could be achieved with only two pulses, but restricting the parameters so that the transfer is perfect for the state with no atoms in $|1\rangle$ initially, we obtained a numerical maximum of the transfer fidelity for the state with one atom in $|1\rangle$ initially of 0.7337. The three pulse scheme is hence necessary.

We note here the convenience of using an intermediate state, which has double occupancy of a single atomic level and hence does not belong to the class of hard core boson states. We imagine, that further exploration of the ladders of occupation number eigenstates may lead to a variety of schemes for efficient state synthesis and manipulation.

\subsection{Braiding}

Excitations of topological states correspond to anyonic quasiparticles, and we next describe a method to demonstrate the phase factor that arises when two such anyons are created and one is moved around the other one. Note that our qubits are not localized in space, but we may nevertheless regard the process as braiding due to the analogy between our system and a plaquette in an optical lattice. As in \cite{paredes}, we consider the state
\begin{equation}
|\Box\rangle=\frac{1}{\sqrt{2}}(|1010\rangle+|0101\rangle),
\end{equation}
which is a minimum instance of a string-net condensate. If two Rydberg levels $|r\rangle$ and $|r'\rangle$ are available and the presence of a single atom in $|r\rangle$ shifts the energy of $|r'\rangle$ significantly, this state can be prepared from $|1010\rangle$ as follows. We first apply a Raman pulse between $|1\rangle$ and $|2\rangle$ to obtain the state $(|1010\rangle+|0110\rangle)/\sqrt{2}$. We then apply (i) a $\pi$ pulse between $|1\rangle$ and $|r\rangle$, (ii) a $\pi$ pulse between $|3\rangle$ and $|r'\rangle$, (iii) a $\pi$ pulse between $|4\rangle$ and $|r'\rangle$, and (iv) a $\pi$ pulse between $|1\rangle$ and $|r\rangle$. This sequence moves the atom in $|3\rangle$ to the level $|4\rangle$ if $|1\rangle$ is unoccupied and hence produces the desired state. With only one Rydberg level available, we can use the procedure in Fig.~\ref{phiplus} to prepare the state $(|2000\rangle+|0101\rangle)/\sqrt{2}$ and then apply a $\pi/\sqrt{2}$ pulse between $|1\rangle$ and $|r\rangle$ followed by a $\pi$ pulse between $|3\rangle$ and $|r\rangle$ to transfer a single atom from $|1\rangle$ to $|3\rangle$ in the first term. The reasoning behind this procedure is that we need to include a pulse sequence, which utilizes the dependence of the Rabi frequency for a transition between a register state and the Rydberg level on the total number of atoms occupying the two levels in order to obtain a qubit controlled nonlinear effect with only one Rydberg level. On the other hand, we can not transform, for instance, $|2000\rangle$ into $|1010\rangle$ or \textit{vice versa} only by use of Raman transitions, and we thus use the Rydberg level a second time to achieve the desired state.

If we denote the Pauli spin operators of the $i$th qubit by $(\sigma_i^x,\sigma_i^y,\sigma_i^z)$, we obtain a pair of so-called chargelike quasiparticle excitations if we apply $\sigma_1^x$, $\sigma_2^x$, $\sigma_3^x$, or $\sigma_4^x$ to $|\Box\rangle$. A pair of fluxlike quasiparticle excitations, on the other hand, is obtained if we apply $\sigma_1^z$, $\sigma_2^z$, $\sigma_3^z$, or $\sigma_4^z$ to $|\Box\rangle$ \cite{paredes}. These excitations have anyonic character, \textit{i.e.}, they accumulate non-trivial phase factors when moved around each other. In \cite{paredes}, it is thus proposed to move a chargelike quasiparticle excitation around the plaquette by subsequent application of $\sigma_i^x$ operators and study the dependence of the outcome on the prior excitation of a pair of fluxlike quasiparticles, of which one is located at the center of the plaquette.

\begin{figure}
\includegraphics[width=\columnwidth]{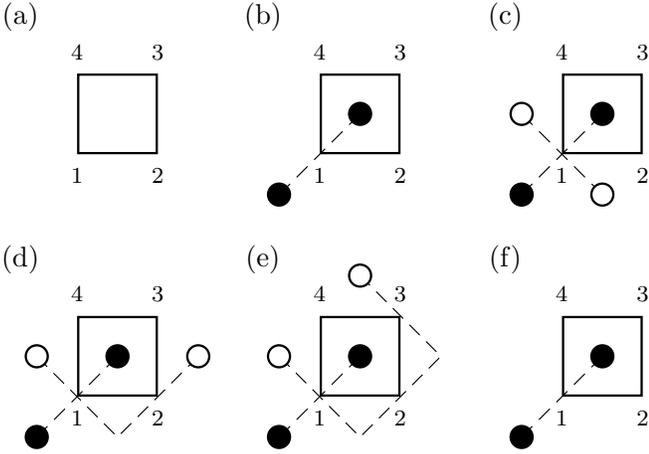}
\caption{Braiding of a chargelike quasiparticle around a fluxlike quasiparticle. (a) The plaquette seen from above. (b) Application of $\sigma_1^z$ creates a pair of fluxlike quasiparticles (black circles). (c) $\sigma_1^x$ creates a pair of chargelike quasiparticles (white circles). (d-e) $\sigma_2^x$ and $\sigma_3^x$ move one of the chargelike quasiparticles. (f) $\sigma_4^x$ annihilates the two chargelike quasiparticles. Finally, the two fluxlike quasiparticles are annihilated by applying $\sigma_1^z$, and we return to (a) except for a phase change of $\pi$.}\label{braid}
\end{figure}

Specifically, application of the sequence of operators $\sigma_4^x\sigma_3^x\sigma_2^x\sigma_1^x$ creates a pair of chargelike quasiparticle excitations, moves one of them around the plaquette, and annihilates the pair again, which altogether transforms $|\Box\rangle$ into $|\Box\rangle$. On the other hand, if we first apply $\sigma_1^z$ to create a pair of fluxlike quasiparticle excitations $\sigma_1^z|\Box\rangle$ and then apply the  $\sigma_4^x\sigma_3^x\sigma_2^x\sigma_1^x$ sequence, we obtain $-\sigma_1^z|\Box\rangle$ with the desired change in sign. This is illustrated in Fig.~\ref{braid}. To observe the sign change experimentally, one can add an extra control qubit and only apply $\sigma_1^z$ if the control qubit is in the state $|0\rangle$. Explicitly, the recipe is as follows. (i) Start from the state
\begin{equation}
(|0\rangle+|1\rangle)\otimes(|1010\rangle+|0101\rangle)/2.
\end{equation}
where the first factor is the state of the control qubit and the second is the state of the four register qubits. (ii) Apply $\sigma_1^z$ if and only if the control qubit is in the state $|0\rangle$ to obtain
\begin{equation} |0\rangle\otimes(|1010\rangle-|0101\rangle)/2+|1\rangle\otimes(|1010\rangle+|0101\rangle)/2.
\end{equation}
(iii) Apply the operator $\sigma_4^x\sigma_3^x\sigma_2^x\sigma_1^x$, which changes the sign of the first term but not of the second, i.e.,
\begin{equation} -|0\rangle\otimes(|1010\rangle-|0101\rangle)/2+|1\rangle\otimes(|1010\rangle+|0101\rangle)/2. \end{equation}
(iv) Apply $\sigma_1^z$ if and only if the control qubit is in the state $|0\rangle$ to obtain
\begin{equation}
(-|0\rangle+|1\rangle)\otimes(|1010\rangle+|0101\rangle)/2.
\end{equation}
(v) Measure the state of the control qubit in the basis $(\pm|0\rangle+|1\rangle)/\sqrt{2}$. Due to the change of sign of the first term in step (iii), the control qubit should be found in the state $(-|0\rangle+|1\rangle)/\sqrt{2}$ rather than $(|0\rangle+|1\rangle)/\sqrt{2}$.

\begin{figure}
\includegraphics[width=\columnwidth]{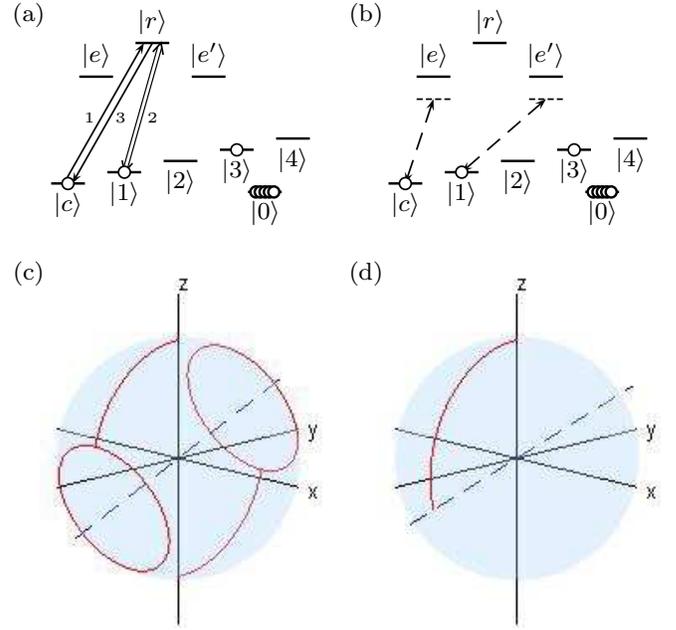}
\caption{(Color online) Controlled phase gate with a single Rydberg level. (a) A $\pi$ pulse (1) transfers the population in $|c\rangle$ to $|r\rangle$, a composite pulse (2) is applied between $|r\rangle$ and $|1\rangle$, which imprints phases but does not alter the populations of the levels, and finally a second $\pi$ pulse (3) transfers the population in $|r\rangle$ back to $|c\rangle$. (For simplicity, only the term $|1\rangle|1010\rangle$ is shown.) (b) Appropriate single qubit phase shifts are applied by coupling $|c\rangle$ and $|1\rangle$ off-resonantly to excited levels. (c-d) The composite pulse sequence illustrated on the Bloch sphere for the case of a total population in $|c\rangle$ and $|1\rangle$ of one atom (c) and a total population in $|c\rangle$ and $|1\rangle$ of two atoms (d). Note that the Rabi frequency is enhanced by a factor of $\sqrt{2}$ in the latter case compared to the former case. See text for further details.}\label{cphase}
\end{figure}

The controlled phase gate required in steps (ii) and (iv) can be implemented (up to an overall unimportant minus sign) by the following pulse sequence if two Rydberg levels are available: (i) A $\pi$ pulse between $|r\rangle$ and the level $|c\rangle$ encoding the state of the control qubit, (ii) a $2\pi$ pulse between $|1\rangle$ and $|r'\rangle$, and (iii) a $\pi$ pulse between $|c\rangle$ and $|r\rangle$. If only one Rydberg level is available, we propose to use the pulse sequence illustrated in Fig.~\ref{cphase}. The composite pulse (2) in Fig.~\ref{cphase}(a) consists of three pulses with the following parameters, where $\phi_i$ is the phase of pulse $i$, $\Omega_i$ is the norm of the Rabi frequency of pulse $i$, $\Delta_i=\omega_i-(E_r-E_1)/\hbar$, $\omega_i$ is the frequency of pulse $i$, $(E_r-E_1)$ is the energy of the Rydberg level relative to the energy of level $|1\rangle$, and $t$ is the duration of each of the pulses:
\begin{align}\label{puls1}
\phi_1&=0, & \Omega_1 t&=\Omega t, & \Delta_1 t&=0,
\end{align}
\begin{multline}\label{puls2}
\phi_2=\pi/2,\quad\Omega_2 t=\frac{2\pi\sin(\sqrt{2}\Omega t)} {\sqrt{1+\cos^2(\sqrt{2}\Omega t)}},\\
\Delta_2 t=-\frac{2\sqrt{2}\pi\cos(\sqrt{2}\Omega t)}{\sqrt{1+\cos^2(\sqrt{2}\Omega t)}},
\end{multline}
\begin{align}\label{puls3}
\phi_3&=\pi, & \Omega_3 t&=\Omega t, & \Delta_3 t&=0.
\end{align}

Let us first consider the case of one atom in $|c\rangle$ and no atom in $|1\rangle$. The starting point for the composite pulse is then the state with one atom in $|r\rangle$ and no atom in $|1\rangle$, which corresponds to the north pole of the Bloch sphere. As shown in Fig.~\ref{cphase}(c), pulse 1 rotates the state an angle $\Omega t$ around the $x$-axis, pulse 2 rotates the state an angle $\sqrt{(\Omega_2 t)^2+(\Delta_2 t)^2}=2\pi$ around the dashed axis in the figure, and pulse 3 rotates the state an angle $\Omega t$ around the negative $x$-axis. As a result, we return to the initial state, except that the phase has changed by
\begin{equation}
\Delta\theta_{10}=\pi-\frac{\sqrt{2}\pi\cos(\sqrt{2}\Omega t)}{\sqrt{1+\cos^2(\sqrt{2}\Omega t)}}.
\end{equation}
For the case with no atom in $|c\rangle$ and one atom in $|1\rangle$, we start from the south pole but otherwise the picture is the same, and the phase of the state is again changed by $\Delta\theta_{01}=\Delta\theta_{10}$.

When $|c\rangle$ and $|1\rangle$ are both populated by one atom, we start from the north pole, but the Rabi frequency is enhanced by a factor of $\sqrt{2}$ compared to the values stated in Eqs.~\eqref{puls1}, \eqref{puls2}, and \eqref{puls3}. As a consequence, the second pulse is no longer a $2\pi$ pulse. The parameters are, however, chosen such that the rotation axis of the second pulse (the dashed axis in the figure) passes through the point representing the state after the first pulse, and we thus again end up on the north pole after the pulse sequence is completed. In this case the phase shift of the state evaluates to
\begin{equation}
\Delta\theta_{11}=\frac{\sqrt{2}\pi(1-\cos(\sqrt{2}\Omega t))}{\sqrt{1+\cos^2(\sqrt{2}\Omega t)}}.
\end{equation}
If $|c\rangle$ and $|1\rangle$ are both unoccupied, the pulses have no effect, and the phase change is $\Delta\theta_{00}=0$.

The phase changes obtained through the pulses shown in Fig.~\ref{cphase}(b) add $0$ to $\Delta\theta_{00}$, $\sqrt{2}\pi\cos(\sqrt{2}\Omega t)/(1+\cos^2(\sqrt{2}\Omega t))^{1/2}$ to $\Delta\theta_{10}$ and $\Delta\theta_{01}$, and $2\sqrt{2}\pi\cos(\sqrt{2}\Omega t)/(1+\cos^2(\sqrt{2}\Omega t))^{1/2}$ to $\Delta\theta_{11}$. Choosing $\Omega t$ according to
\begin{equation}
\cos(\sqrt{2}\Omega t)=\sqrt{3}-2,
\end{equation}
the total phase shifts then take the values $0$, $\pi$, $\pi$, and $\pi$, respectively.

For the encoding illustrated in Fig.~\ref{two}, we note that it is easier to apply $\sigma_i^z$ operations than $\sigma_i^x$ operations, since the former do not involve application of the Rydberg blockade, and it is hence easier to move a fluxlike quasiparticle around a chargelike quasiparticle. The price to pay, however, is the difficulty in preparing the ground state of a plaquette, which supports generation of a chargelike quasiparticle located at the center of the plaquette.

\subsection{Spinons}

As a final example, we consider spinon excitations of RVB states. The states under consideration are $(|0100\rangle-|0010\rangle)/\sqrt{2}$, $(-|0100\rangle+|0001\rangle)/\sqrt{2}$, and $(|0010\rangle-|0001\rangle)/\sqrt{2}$, which represent spinons at positions 1 and 4, 1 and 3, and 1 and 2, respectively. All these states are easy to generate in our system because the Rydberg blockade is only needed to move precisely one atom from the reservoir state to one of the states representing qubits. In the basis $(|0100\rangle,|0010\rangle,|0001\rangle)$, we can write the states as $(1,-1,0)^T/\sqrt{2}$, $(-1,0,1)^T/\sqrt{2}$, and $(0,1,-1)^T/\sqrt{2}$, respectively.

An interesting property of the states is that they only span a two-dimensional space. This space is orthogonal to $(|0100\rangle+|0010\rangle+|0001\rangle)/\sqrt{3}$, which can be experimentally demonstrated by application of two Raman pulses and a measurement. We can, for instance, choose the Raman pulses to apply the transformation
\begin{equation}
\left[\begin{array}{ccc}
\sqrt{\frac{1}{3}}&\sqrt{\frac{2}{3}}&0\\
-\sqrt{\frac{2}{3}}&\sqrt{\frac{1}{3}}&0\\
0&0&1
\end{array}\right]
\left[\begin{array}{ccc}
1&0&0\\
0&\sqrt{\frac{1}{2}}&\sqrt{\frac{1}{2}}\\
0&-\sqrt{\frac{1}{2}}&\sqrt{\frac{1}{2}}
\end{array}\right]
\end{equation}
in the basis $(|0100\rangle,|0010\rangle,|0001\rangle)$. This transforms our three states $(1,-1,0)^T/\sqrt{2}$, $(-1,0,1)^T/\sqrt{2}$, and $(0,1,-1)^T/\sqrt{2}$ into $(0,-\sqrt{3},1)^T/2$, $(0,\sqrt{3},1)^T/2$, and $(0,0,-1)^T$, respectively. A measurement of the state of the qubit represented by level $|2\rangle$ should thus always result in the outcome `0', while $(1,1,1)^T/\sqrt{3}$ is transformed into $(1,0,0)^T$, which populates level $|2\rangle$ with certainty. Spinon excitations of the above kind with spin singlet pairs of two out of three possible spins (and a random state of the uncoupled spin) have recently occurred as candidate states for reference frame-free quantum key distribution \cite{englert,phot}.

\section{Conclusion}\label{conclusion}

In conclusion, we have proposed to prepare and demonstrate basic properties of minimum instances of topological matter in collective excitations of a small atomic ensemble. State manipulation is achieved by application of the Rydberg blockade mechanism and Raman transitions between different atomic levels. In particular, a number of transformations that may be difficult to carry out on individually selected atoms and pairs of atoms in an optical lattice can be achieved in our system simply by application of uniform Raman pulses on the entire sample. We have provided a general criterion to determine whether two states with two excitations each can be transformed into each other only by application of Raman processes, and we have used this result to derive simple schemes to prepare and manipulate a number of topological states. Experiments on Rydberg blockade have recently demonstrated controlled two-qubit gates between two atoms \cite{gategrangier,gatesaffman}, and we hope that the present work will provide inspiration for future experiments demonstrating few qubit minimum instances of topological matter. Using more internal levels of the atoms and possibly more atomic ensembles, our scheme can also be extended to generation of topological states of a larger number of qubits. As discussed further in \cite{holmium}, Holmium atoms are particularly suitable for this purpose, since they have 128 hyperfine ground state levels.

Following the proposal in \cite{phot}, the states $|b_1b_2\ldots b_N\rangle$ prepared in collective states of the order of a thousand atoms can be effectively converted into $N$ well collimated wave packets with $b_i$ photons in the $i$th packet. Our medium may hence serve as a deterministic source of multi-mode field states with a potential of carrying the topological quantum correlations over long distances. We believe that with the potentially high rate of successful generation, the ability to readily create states of higher complexity than a single plaquette, and also the possibility to either absorb the photon states again in another ensemble or interfere them with the deterministic output from another atomic ensemble, the atomic ensemble collective coding is a serious candidate for a quantum repeater for quantum communication protocols with toric code qubit protection \cite{Hollenberg}.

Finally, we recall that within the collective encoding scheme, we use the Rydberg blockade to restrict the occupation of the internal register states to the qubit values zero and unity. Higher population of these states can also be controlled corresponding to higher spin systems and oscillators, for which the plaquette systems offer a wide range of interesting topological effects -- even in the case of coherent states of oscillators, corresponding to classical fields in optics \cite{taylor}, and to entirely non-interacting particles driven by independent Raman transitions in our atomic ensembles.

\begin{acknowledgments}
This work was supported by ARO-DTO Grant No. 47949PHQC and the European Union Integrated Project AQUTE.
\end{acknowledgments}


\begin{thebibliography}{99}

\bibitem{Feynman} R. Feynman, Int. J. Theor. Phys. \textbf{21}, 467 (1982).

\bibitem{Lloyd} S. Lloyd, Science \textbf{273}, 1073 (1996).

\bibitem{Cirac_simulator} E. Jan\'{e}, G. Vidal, W. D\"{u}r, P. Zoller, J. Cirac, Quant. Inf. Comput. \textbf{3}, 15 (2003).

\bibitem{Lewenstein} M. Lewenstein, A. Sanpera, V. Ahufinger, B. Damski, A. Sen, and U. Sen, Adv. Phys. \textbf{56}, 243 (2007).

\bibitem{kitaev} A. Yu. Kitaev, Ann. Phys. \textbf{303}, 2 (2003).

\bibitem{pachos} J. K. Pachos, Int. J. Quant. Inf., \textbf{4}, 947 (2006).

\bibitem{whaley} C. M. Herdman, K. C. Young, V. W. Scarola, M. Sarovar, and K. B. Whaley, Phys. Rev. Lett. \textbf{104}, 230501 (2010).

\bibitem{nayak} C. Nayak, S. H. Simon, A. Steran, M. Freedman, and S. D. Sarmam, Rev. Mod. Phys. \textbf{80}, 1083 (2008).

\bibitem{DiVincenzo2009} D. P. DiVincenzo, Phys. Scr. \textbf{T137}, 014020 (2009).

\bibitem{Hollenberg} A. G. Fowler, D. S. Wang, and L. C. L. Hollenberg, arXiv:1004.0255.

\bibitem{Jaksch} D. Jaksch, C. Bruder, J. I. Cirac, C. W. Gardiner, and P. Zoller, Phys. Rev. Lett. \textbf{81}, 3108 (1998).

\bibitem{Greiner} M. Greiner, O. Mandel, T. Esslinger, T. W. H\"{a}nsch, and I. Bloch, Nature \textbf{415}, 39 (2002).

\bibitem{IsingAK} A. S\o rensen and K. M\o lmer, Phys. Rev. Lett. \textbf{83}, 2274 (1999).

\bibitem{lattice-gauge} H. P. B\"{u}chler, M. Hermele, S. D. Huber, M. P. A. Fisher, and P. Zoller, Phys. Rev. Lett. \textbf{95}, 040402 (2005).

\bibitem{weimer} H. Weimer, M. M\"uller, I. Lesanovsky, P. Zoller, H. P. B\"uchler, Nature Phys. \textbf{6}, 382 (2010).

\bibitem{paredes} B. Paredes and I. Bloch, Phys. Rev. A {\bf77}, 023603 (2008).

\bibitem{brion} E. Brion, K. M\o lmer, and M. Saffman, Phys. Rev. Lett. {\bf99}, 260501 (2007).

\bibitem{pan} C.-Y. Lu, W.-B. Gao, O. G\"{u}hne, X.-Q. Zhou, Z.-B. Chen, and J.-W. Pan, Phys. Rev. Lett. \textbf{102}, 030502 (2009).

\bibitem{weinfurter} J. K. Pachos, W. Wieczorek, C. Schmid, N. Kiese, R. Pohlner, and H. Weinfurter, New J. Phys. \textbf{11}, 083010 (2009).

\bibitem{tordrup} K. Tordrup and K. M\o lmer, Phys. Rev. A \textbf{77}, 020301(R) (2008).

\bibitem{jaksch} D. Jaksch, J. I. Cirac, P. Zoller, S. L. Rolston, R. C\^{o}t\'{e}, and M. D. Lukin, Phys. Rev. Lett. {\bf 85}, 2208 (2000).

\bibitem{lukin} M. D. Lukin, M. Fleischhauer, R. C\^{o}t\'{e}, L. M. Duan, D. Jaksch, J. I. Cirac, and P. Zoller, Phys. Rev. Lett. {\bf 87}, 037901 (2001).

\bibitem{Rydbergreview} M. Saffman, T. G. Walker, and K. M{\o}lmer, Rev. Mod. Phys. {\bf82}, 2313 (2010).

\bibitem{brionerror} E. Brion, L. H. Pedersen, M. Saffman, and K. M{\o}lmer, Phys. Rev. Lett. {\bf100}, 110506 (2008).

\bibitem{Takagi} R. A. Horn and C. R. Johnson, Matrix analysis, Cambridge University Press (1990).

\bibitem{englert} G. Tabia and B.-G. Englert, arXiv:0910.5375.

\bibitem{holmium} M. Saffman and K. M{\o}lmer, Phys. Rev. A {\bf78}, 012336 (2008).

\bibitem{phot} A. E. B. Nielsen and K. M\o lmer, Phys. Rev. A \textbf{81}, 043822 (2010).

\bibitem{gategrangier} T. Wilk, A. Ga\"etan, C. Evellin, J. Wolters, Y. Miroshnychenko, P. Grangier, and A. Browaeys, Phys. Rev. Lett. {\bf104}, 010502 (2010).

\bibitem{gatesaffman} L. Isenhower, E. Urban, X. L. Zhang, A. T. Gill, T. Henage, T. A. Johnson, T. G. Walker, and M. Saffman, Phys. Rev. Lett. {\bf104}, 010503 (2010).

\bibitem{taylor} M. Hafezi, private communication.


\end{thebibliography}
\end{document}